\documentstyle[12pt]{article}
\newcommand{\be}{\begin{equation}}
\newcommand{\ee}{\end{equation}}
\newcommand{\ba}{\begin{eqnarray}}
\newcommand{\ea}{\end{eqnarray}}
\renewcommand{\thefootnote}{\fnsymbol{footnote}}
\begin{document}
\begin{center}
{\Large 
Systems with Higher-Order Shape Invariance: 
Spectral and Algebraic Properties}\\

\vspace{1cm}

{\large A. Andrianov} $^{a,b,}$\footnote{{\it E-mail:} 
andrian@snoopy.phys.spbu.ru; \quad andrian@ecm.ub.es}, 
{\large F. Cannata} $^{c,}$\footnote{{\it E-mail:} cannata@bo.infn.it},
{\large M. Ioffe} $^{a,}$\footnote{{\it E-mail:} ioffe@snoopy.phys.spbu.ru}, 
{\large D. Nishnianidze} $^{a,}$\footnote{{\it E-mail:} 
david@heps.phys.spbu.ru}\\
$^a$ \  {\small\it Department of Theoretical
Physics, University of Sankt-Petersburg,198904 Sankt-Petersburg, Russia}\\
$^b$ \ {\small\it Departament d'Estructura i Constituents de la Mat\`eria, 
 Universitat de Barcelona, Diagonal, 647, 08028 Barcelona, Spain}\\
$^c$ \ {\small\it Dipartimento di Fisica and INFN, Via Irnerio 46,
40126 Bologna, Italy}
\end{center}

\vspace{1cm}

\noindent 
{\bf Abstract}

\bigskip

{\small 
We study 
a complex intertwining relation of second order 
for Schr\"odinger operators and
construct third order symmetry operators 
for them.
A modification of this approach
leads to a higher order
shape invariance. We analyze with particular attention irreducible second order
Darboux transformations which together with the first order act as building 
blocks. 
For the third order shape-invariance
irreducible Darboux transformations entail only
one sequence of equidistant levels while for the reducible case 
the structure 
consists of up to three infinite sequences
of equidistant levels and, in some cases, singlets or doublets
of isolated levels. }

\bigskip

 
\bigskip
\setcounter{footnote}{0}
\renewcommand{\thefootnote}{\arabic{footnote}}
\section*{\large\bf 1.\quad Introduction}
\hspace*{2ex} 
During the last two decades Supersymmetric Quantum Mechanics became
an important and popular tool in establishing connections between a variety
of isospectral quantum systems \cite{junker}. It gives new insight into
the problem of
spectral equivalence of Hamiltonians, which historically was 
developed as Factorization Method of Schr\"odinger in Quantum Mechanics 
\cite{infeld}
and as Darboux-Crum transformations in Mathematical Physics \cite{matveev}. 

In standard 1-dim SUSY QM \cite{witten} two almost isospectral Hamiltonians 
$H_1$ and $H_2$ are 
combined into a single (matrix) Schr\"odinger operator in a Hilbert 
superspace. The two SUSY partners of a Superhamiltonian are intertwined
by supercharges $Q^{\pm}$
\be
H_1 Q^+ = Q^+ H_2 ; \quad Q^-H_1 = H_2Q^- ;\quad 
Q^-\equiv (Q^+)^{\dagger} \label{intertw} 
\ee
which indeed generate Darboux transformations. 
The superhamiltonian and the two supercharges form the so called SUSY algebra.
Such approach appeared to be very useful in studying a variety of 
different kinds of quantum systems. 
A very fruitful approach is based on generalizations of one specific part 
of the SUSY algebra, namely the 
invariance of the Superhamiltonian in respect to
supertransformations. For the SUSY partner Hamiltonians this invariance
is expressed by intertwining relations.
 
Below we list some specific classes of problems which were investigated 
in this way: 
1) the class of shape-invariant potentials \cite{gendenstein} with the 
algebraic construction of the spectrum;
2) supercharges $Q^{\pm}$ of higher orders (HSUSY) in derivatives 
\cite{ais}--\cite{hussin} including
irreducible 2-nd order transformations \cite{acdi1} and polynomial 
deformation of SUSY algebra; 
3) intertwining of Hamiltonians $H_{1,2}$ with complex potentials by complex 
supercharges up to second order\cite{complex};
4) construction of integrable
2-dim quantum and classical systems and corresponding symmetry operators
\cite{ain1}; 
5) investigation of systems with matrix potentials \cite{matrix}.

In this paper we investigate some further generalizations of 1-dim
SUSY intertwining relations starting from two hermitian Hamiltonians
intertwined by complex supercharges of second order in derivatives (Section 2).
A system of intertwining relations arises so that the same Hamiltonians
are related at the same time by first and second order Darboux transformations.
The compatibility of these constraints require the potentials to satisfy
a stationary KdV equation \cite{lamb}.
This formulation automatically allows (by elimination of one of the partner
Hamiltonians) for a discussion of  symmetry operators $R_i$ 
of third order for each partner separately
$[H_i, R_i ] = 0$. These symmetry operators 
which commute with the Hamiltonians were studied before
in another approach \cite{nikitin},\cite{doebner}, and 
the corresponding potentials were found.

In Section 3 a further generalization of the previous relations 
for the symmetry 
operators is obtained by the ladder equation 
$[H, a ] = 2\lambda a .$
This approach reflects in part results obtained in \cite{veselov}
in their construction of dressing chains in connection with integrable
bi-hamiltonian systems
and in \cite{sukhatme} in the context of numerical illustrations of 
(cyclic) shape-invariance
preserving standard shape invariance in each step.
However our approach also leads to new findings concerning 
explicitly with the role of 
irreducible second order Darboux transformations \cite{acdi1}.
We will focus attention on the non-trivial dependence of the spectrum
on the relations between particular parameters of the system, illustrating
in detail the particular cases of second and third orders. 
We find for the third order different type of spectra involving up to three
sequences of equispaced levels while the potentials are not strictly
harmonic. Conditions for which 
one of these sequences 
is shrinked to one or two levels
are investigated via the study of zero modes of the generalized creation
operator. We stress that these zero modes are allowed only because the creation
operators are here of higher order. 

In the Conclusion (Section 4) we emphasize the original aspects of
our investigation in contradistinction with results in the literature.

\section*{\large\bf 2.\quad Complex supercharges, symmetry operators
of third order and KdV equation}
\hspace*{2ex} 

In this Section we introduce supercharges with complex superpotentials
intertwining two hermitian partner Hamiltonians (\ref{intertw}), where:
\be
H_i = - \partial^2 + V_i(x); \quad \partial\equiv \frac{d}{dx}. \label{ham}
\ee 
Since complex supercharges of first order do not lead to a genuine extension
we start from:
\ba
Q^+ &=& M^+ + i \tilde q^+ ;  \label{Q^+}\\
M^+ &\equiv & \partial^2 - 2f(x)\partial + b(x) ; \label{M^+}\\
\tilde q^+ &\equiv & - 2\tilde f(x)\partial + \tilde b(x) . \label{tildeq} 
\ea
As a consequence of (\ref{intertw}) we obtain that $\tilde f(x)=Const$
and one can rescale the first order $\tilde q^+ \equiv -2\tilde f q^+$
with $q^+$ of standard structure:
\be
q^+ = \partial + W(x); \qquad W(x) \equiv - \frac{\tilde b(x)}{2\tilde f}.
\label{q^+}
\ee 
In terms of the second order $M^+$ and the first order $q^+$ (\ref{intertw})
amounts to a system of two intertwining relations for the same partner
Hamiltonians:
\ba
H_1 M^+ &=& M^+ H_2 ;  \label{intM} \\
H_1 q^+ &=& q^+ H_2 .  \label{intq} 
\ea
Imposing (\ref{intM}) leads \cite{acdi1} to:
\ba
V_{1,2}(x) &=& \mp2f'(x) + f^2(x) + \frac{f''(x)}{2f(x)} - 
\frac{f^{\prime 2}(x)}{4f^2(x)} - \frac{d}{4f^2(x)} + \alpha ;\label{acdi}\\
b(x) &=& -f'(x) + f^2(x) - \frac{f''(x)}{2f(x)} + 
\frac{f^{\prime 2}(x)}{4f^2(x)} + \frac{d}{4f^2(x)} \label{b}
\ea
with arbitrary real constants $d$ and $\alpha .$ 
Let us remind that the sign of $d$ in (\ref{acdi}), (\ref{b}) 
is crucial to characterize the so called \cite{acdi1} reducible
$(d\leq 0)$ and irreducible $(d>0)$ second order transformation 
\footnote{We remind that reducible second order operator $M^+$ can be written
as a product $(\partial + W_1(x))(\partial + W_2(x))$ where the real 
superpotentials $W_1(x), W_2(x)$ satisfy a gluing condition 
$ - W_1'(x) + W_1^2(x)= +W'_2(x) + W^2_2(x) +\sqrt{-d}$
(see details in \cite{acdi1}).}. 

Similarly, (\ref{intq})
gives the standard 
\be
V_{1,2}(x) = \pm W'(x) + W^2(x) ,  \label{beta}
\ee
where an additive constant is set equal to zero.

Consistency requires that
\be
W(x) = -2f(x) + \gamma ; \quad \gamma=const   \label{gamma}
\ee
and that $f(x)$ satisfies a nonlinear second order differential equation
which can be integrated to:
\be
f^{\prime 2}(x) = 4f^4(x) - 8\gamma f^3(x) - 
4 (\alpha - \gamma^2)f^2(x) + \beta f(x) - d ; \quad 
\beta = const. \label{delta}
\ee
Though (\ref{delta}) can be solved for $x=x(f),$
we will not make use of this solution.

It is known \cite{ain1}, \cite{matrix}, that for arbitrary intertwining
relation for hermitian Hamiltonians one can construct symmetry operators
of even order in derivatives which  can be more or less trivial
in 1-dim scalar case but not for 2-dim or matrix case.
Since now we deal with two intertwining relations (\ref{intM}), (\ref{intq}), 
we can provide non-hermitian symmetry operators $R_1=M^+q^-, R_2=
M^-q^+$ of third order, to which of course one can add an arbitrary 
constant:
\ba
H_1 M^+q^- &=& M^+q^- H_1 ; \label{Mq1} \\
H_2 M^-q^+ &=& M^-q^+ H_2 . \label{Mq2}
\ea
One finds that the hermitian part of $R_i$ is 
trivially proportional to
$H_i$ but its antihermitian part $A_i$ is more involved:
\be
A_1 \simeq -\partial^3 + \bigl( 6f^3-3f'-6\gamma f-\alpha +\gamma^2 \bigr)
\cdot\partial +
\bigl( -12f^3+18\gamma f^2+6(\alpha -\gamma^2)f+6ff'-3\gamma f' \bigr)   
\label{anti}
\ee
with similar expression for $A_2.$

A systematic investigation of symmetry operators of higher order in
derivatives was performed for 1-dim Schr\"odinger operator in \cite{nikitin},
following the basic strategy of writing an operator 
polynomial in derivatives and imposing that it commutes with the Hamiltonian.
By this procedure one obtains that third order symmetry operators 
exist if the potential satisfies the stationary version of KdV equation 
\cite{lamb}:
\be
V^{\prime\prime}(x) = 3V^2(x) - c_1V(x) + c_0    \label{kdv}
\ee 
where $c_0,c_1$ are real constants. One can check that symmetry operator
(\ref{anti}) with $f(x)$ satisfying (\ref{delta}) can be rewritten as
\be
A_1 \simeq \partial^3 + \bigl( -\frac{3}{2} V_1(x) + 
\frac{c_1}{4} \bigr)\cdot\partial - \frac{3}{4} V^{\prime}_1(x) 
\label{anti2}
\ee
which indeed coincides with the expression obtained in
\cite{nikitin} for potentials $V_1(x)$ for which (\ref{kdv})
holds.

In general, the solution of (\ref{kdv}) can be expressed \cite{lamb} 
in terms of periodic Jacobi elliptic function:
\be
V(x) = \kappa_1 sn^2(\kappa_2 x; k) + \kappa_3 , \label{sn} 
\ee
where the constants $\kappa_i$ are related \cite{lamb} to the constants $c_i$ 
of (\ref{kdv}).
We focus attention now to the interplay between existence of a symmetry
operator and the "non-degeneracy" of normalizable solutions in 
1-dim Quantum Mechanics. For this
purpose we restrict ourselves to the period $k=1,$ which corresponds to
$sn(\kappa_2 x; k=1) = th (\kappa_2 x).$ It is straightforward to check
explicitly that the only bound state of the associated potential $V_1(x)$ 
is an eigenstate of $A_1.$ More generally, as suggested by a fundamental 
theorem in Mechanics, one can convince oneself that the operator 
$A_1^2$ is a third
order polynomial with constant coefficients in $H_1$ for $V_1(x)$ 
satisfying (\ref{kdv}). 

As a final comment let us mention how one can extend the third order
results to higher orders, for example to the fifth order symmetry operators.
Imposing directly the commutation between the fifth order symmetry operator
and the Hamiltonian, one obtains as a consistency relation that $V(x)$
has to satisfy a higher KdV equation \cite{lamb} and that this symmetry 
operator
can be written as a product $H A_1$ with $A_1$ given in (\ref{anti2}).
One can notice that if $V(x)$ satisfies (\ref{kdv}) it will also
automatically satisfy the higher KdV (but not vice versa).

\section*{\large\bf 3.\quad Intertwining with shift and higher order 
shape-invariance}
\hspace*{3ex}
For illustration of the construction let us start by modifying 
equations (\ref{intM}), (\ref{intq}):
\ba
H_1 M^+ &=& M^+ H_2 ;  \label{inttM} \\
H_1 q^+ &=& q^+ (H_2 + 2\lambda) ,  \label{inttq} 
\ea
introducing thereby a shift by the positive constant $2\lambda .$
Eliminating one partner Hamiltonian (for example, $H_2 )$ and
denoting product operators $a^+ \equiv q^+M^-$ and
$a^- \equiv M^+q^-$ we obtain:
\be
H_1 a^+ = a^+ (H_1 + 2\lambda) .  \label{shape}
\ee 
We will call such Hamiltonian as "third order 
shape-invariant". 
One can also work the other way around and start from (\ref{shape})
with an operator $a^+$ of third order represented as a product
$q^+M^-$ and $H_1$ represented by $H_1=q^+q^-$ as is familiar
in SUSY Quantum Mechanics. After one can introduce an auxiliary
Hamiltonian $H_2 + 2\lambda = q^-q^+$ to obtain (\ref{inttM}).

Eq.(\ref{shape}) is a ladder equation where $a^+$ plays the role of
generalized creation operator which provides an excitation energy
of $2\lambda .$ In order to study the spectrum it is crucial to study
zero modes of $a^-$ and $a^+.$ The former describe the lowest lying
levels of the system, and one has to apply recursively 
the operator $a^+$ to them in order to generate the excitation spectrum.
The energies of the zero modes can be obtained by imposing the vanishing
of their norm, which involves the average of the operator product 
$a^+a^- .$ This product can be easily evaluated algebraically because 
\cite{acdi1} 
\be
a^+a^- = q^+M^-M^+q^- = q^+\bigl( (H_2-\alpha)^2 + d \bigr)q^- = 
H_1\bigl( (H_1-\alpha - 2\lambda)^2 + d \bigr). \label{prod1}
\ee
Contrary to the standard harmonic oscillator, one has the possibility 
to have also zero modes of the operator $a^+$ which correspond to a
possible truncation of the tower of excited levels. Arguing as before, 
the relevant operator product is in this case 
\be
a^-a^+ = (H_1+2\lambda)
\bigl((H_1-\alpha )^2+d\bigr) \label{prod2}
\ee
to be averaged over the excited states.

This construction can be generalized to higher orders
taking into account that on general grounds \cite{acdi1} $n-$th
order operators $a^{\pm}$ can always be constructed in terms of products
of $q$ and $M.$
In the general higher order case, (\ref{shape}) gives a 
connection between $H_1$ and $H_1$ plus shift, which is the simplest 
realization of the notion of shape-invariance \cite{gendenstein}. 
Though some properties of the spectrum, like zero modes of $a^{\pm}$,
will depend on the explicit product structure of $a^{\pm},$ 
the excitation spectrum can be 
mainly
obtained algebraically.
So it is possible to study the consequences of (\ref{shape})
without taking into account the specific definition of the operators
$a^{\pm}.$ 

In order to provide additional arguments to the interpretation of 
Eq.(\ref{shape}) as a generalization of shape invariance,
let us start to notice that if $a^{\pm}$ 
would be of first order,
$H_1$ would be identified with harmonic oscillator. If $a^{\pm}$ 
is of second order,
we reobtain the singular harmonic oscillator potential which is also
shape-invariant (see next 
Subsection). 
We stress that if $a^{\pm}$ would be an operator of second order, it is known 
\cite{acdi1} that it is not always possible to write it as a product of
two first operators with real superpotentials: such case is referred
as irreducible. 

\subsection*{\large\bf 3.1.\quad Intertwining with shift and second order 
shape-invariance}
\hspace*{3ex}
In this Subsection to keep notations the same we consider 
(\ref{inttM}), (\ref{inttq}) with $M^+$ temporarily of first order.
We can thus
explore in this Subsection if 
even a first order intertwining operator $M^+$ can now
lead to nontrivial consequences. It is easy to convince oneself that
this is indeed so. The superpotentials which solve (\ref{inttM}), 
(\ref{inttq}) are superpositions of a term growing as $x$ and a 
singular term
like $1/x$ leading to two singular potentials \cite{gango} 
$V_1(x)$ and $V_2(x)$ as follows\footnote{On the half line for 
$ \rho =l+1 $ and suitable choice of boundary conditions for
wave functions \cite{acdi1} potential $ V_{1} $ can be interpreted
as a radial harmonic oscillator in partial wave $ l .$}:
\be
V_1(x) = \frac{\rho(\rho - 1)}{x^2} + \frac{\lambda^2x^2}{4};\qquad
V_2(x) = \frac{\rho(\rho + 1)}{x^2} + \frac{\lambda^2x^2}{4} - \lambda .
\label{sing}
\ee
These potentials are shape-invariant in the standard sense \cite{gendenstein} 
and belong to the class of algebraically solvable models,
because $V_2(x;\rho , \lambda) = V_1(x;\rho + 1 , \lambda) - \lambda .$
 
It is instructive to display the algebraic properties of these
systems. In terms of the product operators $a^+ \equiv q^+M^-$ and
$a^- \equiv M^+q^-$ (we repeat that in this Subsection $a^{\pm}$ are of 
second order) one obtains an algebra suggestive of a generalization
of the standard harmonic oscillator algebra:
\be
[ H_1, a^+ ] = 2\lambda a^+ ; \quad [ a^+, a^- ] = -4\lambda H_1 + const . 
\label{alg}
\ee
Similar results hold for $H_2.$ 

It is now tempting to consider, as explained before, Eqs.(\ref{alg}) by
themselves including the case where the operators
$a^{\pm}$ are irreducible. We know from \cite{acdi1} with
$H_2 \equiv H_1 + 2\lambda$ and setting\footnote{This choice of
the energy scale means \cite{acdi1} that the eigenvalues 
of $ H_{1}$ are bounded from below by $ -\sqrt{-d}.$} \, 
$\alpha\equiv 0$ in (\ref{acdi}) that
\be
a^+a^- = H_1^2 + d ;\quad a^-a^+ = (H_1+2\lambda)^2 + d . \label{both}
\ee

Solving (\ref{shape}) we obtain from (\ref{acdi}) that 
$f(x)=\frac{1}{2}\lambda x$ and 
\be
V_1(x) = \frac{\lambda^2x^2}{4} - (\frac{1}{4} + \frac{d}{\lambda^2})
\frac{1}{x^2} - \lambda . \label{second}
\ee
Evaluating the matrix element of the first of (\ref{both}), 
we get the equation for the $0$-modes of $a^-$: 
$E^{(0)}_{1,2} = \pm\sqrt{-d} .$ For a positive value of $d$
(irreducible case) there are no solutions and physically this is clear 
because the potential (\ref{second}) describes a collapse for $d>0 .$
For $d\leq 0$ the singular part of the potential is less pathological 
and one may have two zero modes for suitable choices of parameters. 
The eigenfunctions of the zero modes of $a^-$ which are also eigenfunctions of
$H_1$ can be found explicitly 
by replacing $\partial^2$ in $a^-$ by $(V_1(x) - E^{(0)}).$  
The asymptotic behavior of the corresponding eigenfunctions
is compatible with the normalizability.    

We expect that for self-adjoint extensions of Hamiltonian $H_1$
both operators $a^+a^-$ and $a^-a^+$ are nonnegative on physical
states. There exist two zero modes of $a^-$ if
$\sqrt{-d} < \lambda .$
Acting with $a^+$ on these two zero modes, one can construct
two sequences of levels with internal spacing given by $2\lambda . $
In the limiting case $d=0$ the two sequences coincide. 
When $ \sqrt{-d} > \lambda , $ one obtains only one zero mode with the 
energy $ E^{(0)}=\sqrt{-d} $ and, therefore, only one sequence
of equispaced levels. The existence of the second zero mode 
$ E^{(0)}=-\sqrt{-d} $ in this case is incompatible with positivity of
the second operator in Eq.(\ref{both}). For  $ \sqrt{-d} = \lambda $ this
second zero mode remains non-normalizable.

\subsection*{\large\bf 3.2.\quad Intertwining with shift and third order 
shape-invariance.}
\hspace*{3ex}From now on we 
will consider the intertwining operator $M^{\pm}$ to be 
a reducible or irreducible operator of 
second order (\ref{M^+}) but $q^{\pm}$ still of first order (\ref{q^+}).
The solutions of (\ref{inttM}) are unchanged in respect to Eqs.(\ref{acdi}), 
(\ref{b}) of Section 2 but Eq.(\ref{inttq}) implies a shift for the
potential $V_2 (x)$ in (\ref{beta}). Thus
the consistency equations for $W(x)$ and $f(x)$ are modified. 
Eq.(\ref{gamma}) becomes:
\be
W(x) \equiv W_{3}(x) = -2f(x) - \lambda x , \label{W}
\ee
where an additional integration constant can be ignored because of 
a shift of $x,$ which fixes the origin of coordinate\footnote{In 
Eq.(\ref{gamma})
there was no such fixing, so the limit $\lambda\to 0$ cannot be taken to
obtain the results of Section 2.}.
The potential $V_1(x)$ can be written from (\ref{beta}) as:
\be
V_1(x) = -2f^{\prime}(x) + 4f^2(x) + 4\lambda xf(x) + \lambda^2x^2 -
\lambda , \label{VV}
\ee
with $f(x)$ satisfying:
\be
f^{\prime\prime} = \frac{f^{\prime 2}(x)}{2f(x)} + 6f^3(x) +
8\lambda xf^2(x) + 2(\lambda^2 x^2 - (\lambda + \alpha))f(x) + 
\frac{d}{2f(x)}.    \label{painleve}
\ee 
The equation (\ref{painleve}) can be transformed by the substitution 
$f(x)\equiv 1/2 \sqrt{\lambda} g(y); \,\, y\equiv \sqrt{\lambda}x$ to 
the Painleve-IV equation \cite{ince}:
\be
g^{\prime\prime} = 
\frac{g^{\prime 2}(y)}{2g(y)} + \frac{3}{2} g^3(y) +
4 yg^2(y) + 2(y^2 - a)g(y) - \frac{b}{2g(y)}.    
           \label{Painleve}
\ee
where 
\be
a\equiv 1 + \frac{\alpha}{\lambda};\quad b\equiv -\frac{4d}{\lambda^2} .
\label{ab}
\ee
This equation has been studied intensively in the 
last years \cite{book} and in the following we will be mainly interested
in asymptotic properties of its solutions which will determine the 
asymptotics of potentials (\ref{VV}) and the normalizability of eigenfunctions.

Concerning the algebra, the only modification in respect to (\ref{alg}) in the
previous Subsection is given by:
\be
[ a^+, a^-] = -2\lambda \bigl( 3H_1^2 -(4\alpha + 2\lambda)H_1 + \alpha^2 + d
\bigr)   
\label{alg2} 
\ee
and similarly for $H_2.$ When the index will not appear explicitly 
we intend to refer to the index $1.$
 
We show how one can derive the spectrum
from (\ref{shape}) and (\ref{alg2}) if normalizable zero modes
of the annihilation operator $a^-$ exist. We stress that this algebraic 
method is very powerful now since the explicit form of the potential
is known only in terms of Painleve transcendents.

The equation for zero modes of $a^-$ reads:
\be
a^- \Psi^{(0)}_k = M^+q^-\Psi^{(0)}_k = 0  \label{zero}
\ee
where $k$ labels the normalizable solutions. 

As explained at the beginning of Section 3, Eq.(\ref{prod1}),
the algebraic equation for eigenvalues $E^{(0)}_k $ is:
\be
E^{(0)}\cdot \bigl[(E^{(0)} - \alpha - 2\lambda)^2 + d\bigr]=0 , 
\label{energy}
\ee
and has at most three real solutions.

For the {\it irreducible} case $(d>0)$ only one zero mode  $E^{(0)}_0=0$ 
exists\footnote{In this case we notice
that since $M^-\Psi^{(0)}_0 \neq 0$ also $H_2$ has the ground state with 
zero eigenvalue.}.  
But it is necessary
to check indeed the normalizability of corresponding zero mode which can be
written (\ref{zero}), (\ref{W}), (\ref{q^+}) in terms of superpotentials:
\be
\Psi^{(0)}_0(x) = exp (\int^x W(z)dz) = exp \bigl(-\frac{1}{2}\lambda x^2 -
2\int^xf(z)dz \bigr) \label{psi}
\ee
where $f(z)$ satisfies (\ref{painleve}) and must have a suitable 
asymptotic behavior in order to make (\ref{psi}) normalizable. 
It has been found \cite{kapaev} that for $b<0$ there are two 
possible 
asymptotic behavior of $g(y)$ at $\pm\infty :-2y/3 $ and $ \,-2y.$ 
It has been demonstrated \cite{kapaev} that one can realize 
the matching between
the same asymptotic value at $+\infty$ and $-\infty .$
The asymptotic values which are compatible with normalizability of
(\ref{psi}) 
is $g(y)\sim -2y/3 ,$ 
or equivalently,
$f(x)\sim -\lambda x/3 .$ 
We notice that the subleading term in 
$f(x)$ have oscillations
$\sim cos(\lambda x^2/\sqrt{3} + o(x)) $ which show up in 
the potential (\ref{VV}) as a modulation of the harmonic term.

 From $[H, a^+]=2\lambda a^+ $ one can deduce that $E^{(n)}=2\lambda n .$ 
The related potential (\ref{VV})
for the increasing asymptotic behavior of $ f(x) $
has asymptotics  $\lambda^2x^2/9 ,$ which differs from conventional
$\lambda^2x^2 .$
We remark that the latter system does not belong to the  families 
of potentials
isospectral to harmonic oscillator \cite{nieto},\cite{roy} 
which have
the same asymptotics but differ from harmonic oscillator for finite $x.$

Evaluating the norm of the excited state $\Psi^{(n+1)}(x)\equiv 
a^+\Psi^{(n)}(x),$ one finds after algebraic manipulations explained in 
the beginning of Section 3 that for 
the irreducible case it cannot vanish. So, the spectrum is not bounded from 
above. Analytic calculation support clearly this result 
because of the asymptotic oscillator-like growth.

For the {\it reducible} case $M^{+}$ with $d\leq 0$ is: 
\be
M^{+}\equiv (\partial + W_{1}(x)) (\partial + W_{2}(x)) \label{aproduct}
\ee
with real superpotentials \cite{acdi1}
\be
W_{1,2}(x)=-f(x)\pm\frac{f'(x)-\sqrt{-d}}{2f(x)} .\label{superpot}
\ee
Eq.(\ref{energy}) has the solution
$E^{(0)}_0=0$ as before and two additional solutions:
\ba
E^{(0)}_-=\alpha +2\lambda - \sqrt{-d} \label{one}\\
E^{(0)}_+=\alpha +2\lambda + \sqrt{-d} \label{two}
\ea 
which correspond to eigenvalues of $H=q^+q^-$ provided (\ref{one}), 
(\ref{two}) are non-negative and that associated eigenfunctions
$ \Psi^{(0)}_{k}(x) $ are 
normalizable. 
Acting on the three zero modes $\Psi^{(0)}_k(x)$ by the creation operator
$a^+,$ one creates the excited states of the system organized 
in three sequences.
According to Eq.(\ref{shape}), within each sequence the levels are 
$2\lambda -$equidistant. 
The non-negativity of the norm 
$ ||a^{\pm}\Psi(x)||^{2} $ for all physical
states $ \Psi(x) $ 
leads to additional necessary conditions for the 
spectrum\footnote{From the algebra (\ref{shape}) - (\ref{prod2})
one can prove that these conditions are sufficient to ensure 
the non-negativity of the norm of any state created by a polynomial
of $ a^{\pm} :\,\,||P(a^{+},a^{-})\Psi(x)||^{2} \geq 0 . $ }.

The eigenfunctions $\Psi^{(0)}_{k}(x),$ annihilated by 
$a^-=M^+q^-=(\partial + W_{1}(x))\cdot (\partial + W_{2}(x))\cdot
(-\partial + W_{3}(x))$ with $ W_{3}(x) $ defined in (\ref{W}),
can be calculated  
explicitly: 
\ba
\Psi^{(0)}_{0}(x) &=& exp\bigl( \int^{x}W_{3}(x')dx' \bigr) ;\nonumber\\
\Psi^{(0)}_{+}(x) &=& (W_{2}(x)-W_{3}(x)) 
exp\bigl( -\int^{x}W_{2}(x')dx' \bigr) ; \label{PPsi}\\
\Psi^{(0)}_{-}(x) &=& 
\biggl( 2\sqrt{-d} + (W_{2}(x)-W_{3}(x))(W_{1}(x)+W_{2}(x)) \biggr)
exp\bigl( -\int^{x}W_{1}(x')dx' \bigr) . \nonumber
\ea
Their normalizability, similarly to
$\Psi^{(0)}_0(x)$ for the irreducible case, 
depends on the choice for the asymptotics of $f(x)$
and particular values of parameters $ \alpha , \lambda $ and $ d . $
In reducible case $b>0$ four possible asymptotics of solutions
of Painleve-IV equation can appear: $g(y)\sim -2y/3,\,-2y,$ and
$\pm\sqrt{b}/2y .$

On the other hand, the operator $ a^{+} $ can also have zero modes
which can be found explicitly:
\ba
\Psi_{1}(x) &=& exp\bigl( \int^{x}W_{1}(x')dx' \bigr) ;\nonumber\\
\Psi_{2}(x) &=& (W_{1}(x)+W_{2}(x)) 
exp\bigl( \int^{x}W_{2}(x')dx' \bigr) ; \label{PPPsi}\\
\Psi_{3}(x) &=& 
\biggl( E^{(0)}_{+} + (W_{1}(x)+W_{2}(x))(W_{2}(x)-W_{3}(x)) \biggr)
exp\bigl( -\int^{x}W_{3}(x')dx' \bigr) . \nonumber
\ea
These eigenfunctions are solutions of the Schr\"odinger equation with
eigenvalues:
\be
E_{1}=\alpha -\sqrt{-d}; \quad E_{2}=\alpha +\sqrt{-d}; \quad 
E_{3}=-2\lambda . \label{Ennergy}
\ee
The consistency of the energy spectrum, i.e. the non-negativity of
operators $a^+a^-$ and $a^-a^+ ,$ rules out the possibility to have
a normalizable zero mode with negative energy 
$ E_{3} $ for nonsingular superpotentials.
Concerning the other zero modes, one 
can conclude only that the total number of zero modes
of {\it both} operators $ a^{-} $ and $ a^{+} $ cannot exceed three,
which follows straightforwardly from the conflicting asymptotics (\ref{PPsi})
and (\ref{PPPsi}).

As the spectrum of Hamiltonian is bounded from below and the operator $a^+$
raises its levels, zero modes of $a^+$ represent an obstruction 
to build an infinite sequence of levels. 
Only one zero mode of $a^+$ may exist together with two zero modes
of $a^- .$ In this case the spectrum consists of one infinite 
sequence of levels and a finite band of levels, both generated
by operators $(a^+)^n .$

Let us proceed to a more detailed analysis of different spectrum patterns.

a) \, Three normalizable zero modes (\ref{PPsi}) of $ a^{-} $ and, 
respectively, three equidistant sequences of levels may arise only for
$ \lambda > \sqrt{-d} $\, if\, $\sqrt{-d}<\alpha <2\lambda -\sqrt{-d}$ \, 
or \,
$\sqrt{-d}-2\lambda \leq\alpha <-\sqrt{-d}.$ The corresponding solutions
of Painleve-IV equation must have the leading asymptotics 
$ f(x)\sim -\lambda x/3 $ with subleading oscillations 
$ cos \bigl(\lambda x^{2}/\sqrt{3} + o(x) \bigr). $

b) \, Two normalizable zero modes of $ a^{-} $ (and two sequences of 
levels) may exist for different solutions of Painleve-IV equation
\footnote{We discuss here only the case of equal asymptotics 
of $f(x)$ at $\pm\infty .$ However we are aware of existence of
solutions with the different asymptotics, which may lead to an 
additional variety of potentials with one or two sequences of energy
levels. The admissible range of parameters and zero mode space is
derived then by intersection of those ones for the particular asymptotics 
described below.}. 
Namely, for $ \lambda >\sqrt{-d} $ the solution with asymptotics 
$ f(x)\sim -\lambda x/3 $ provides the fall off of 
$ \Psi^{(0)}_{-}(x), \, \Psi^{(0)}_{+}(x) $ if
$ \alpha > 2\lambda -\sqrt{-d} $ and of 
$ \Psi^{(0)}_{0}(x), \, \Psi^{(0)}_{+}(x) $ if
$ - 2\lambda -\sqrt{-d}< \alpha <- 2\lambda +\sqrt{-d}. $ 
For $ \lambda >\sqrt{-d} $ the solution with asymptotics 
$ f(x)\sim -\sqrt{-d}/2\lambda x $ generates two sequences of levels 
starting from $\Psi^{(0)}_{0}(x), \, \Psi^{(0)}_{-}(x) $ if 
$ 2\lambda n +\sqrt{-d}<\alpha <2\lambda (n+1)-\sqrt{-d}; \, n=0,1,2... $
For $ \lambda < \sqrt{-d} $ two sequences (from $\Psi^{(0)}_{0}(x)$  
and $\Psi^{(0)}_{+}(x)$) are generated for two possible asymptotics 
$ f(x)\sim\sqrt{-d}/2\lambda x $ and $ f(x)\sim -\lambda x/3 $ if 
$ -2\lambda -\sqrt{-d}<\alpha <-\sqrt{-d} . $

c) \, For $ \alpha \leq -2\lambda -\sqrt{-d} $ and arbitrary positive
$ \lambda $ 
one sequence of equidistant levels will be realized\footnote{Let us
remark that particular value $ \alpha = -2\lambda -\sqrt{-d},$ {\it i.e.}
$ E^{(0)}_{+}=0, E^{(0)}_{-}<0, $
just corresponds to the case when Painleve-IV equation (\ref{Painleve})
has a class of particular solutions which coincide with solutions
of the Riccati equation $ g'(y)=g^{2}(y)+2yg(y)+\sqrt{b}.$ Substituting
this Riccati equation into  Eq.(\ref{VV}) one finds that potential 
is the pure harmonic oscillator: $ V_{1}(x)=\lambda^{2}x^{2}-\lambda .$}
with ground state $ \Psi^{(0)}_{0}(x) $ and one of three asymptotics: 
$ f(x)\sim -\lambda x/3, \,\, \sim\pm\sqrt{-d}/2\lambda x .$ For 
$ f(x)\sim -\lambda x/3 $ only one sequence of levels 
will be realized also for $ \lambda=\sqrt{-d}, \,\,\alpha =\pm\sqrt{-d} $
(starting from $ \Psi^{(0)}_{0}(x) $), for $ \lambda =\sqrt{-d}, \,\,
\alpha >-\sqrt{-d} $ (starting from $ \Psi^{(0)}_{-}(x) $) and for 
$ \lambda <\sqrt{-d},\,\, \alpha\geq -2\lambda -\sqrt{-d} $
(starting from $ \Psi^{(0)}_{+}(x) $). For asymptotic behavior 
$ f(x)\sim -\sqrt{-d}/2\lambda x $ and $ \lambda >\sqrt{-d} $ the 
spectrum consists of one sequence of levels if 
$ |\alpha -2\lambda n|<\sqrt{-d},\,\,n=0,1,2,... $ (starts from 
$ \Psi^{(0)}_{-}(x) $) or if $ \alpha =\sqrt{-d} $ (starts from 
$ \Psi^{(0)}_{0} $). Last, for $ f(x)\sim \sqrt{-d}/2\lambda x $
one obtain one sequence with ground state $ \Psi^{(0)}_{+}$ if 
$ \lambda\leq\sqrt{-d} $ and $ \alpha >-\sqrt{-d} $ or if 
$ \lambda >\sqrt{-d} $ and $ |\alpha -2\lambda n| < \sqrt{-d},\,\,
n=0,1,2,... $

For specific values of parameter $ \alpha $ the existence of 
normalizable zero mode of $ a^{+} $ can truncate one of two
sequences ( with ground state $ \Psi^{(0)}_{0}(x) $ ) in item b).

d) \, One additional singlet state \footnote{
Its existence has been mentioned in \cite{slav}} satisfies the equation
$$ a^{+}\Psi^{(0)}_{0}(x) = a^{-}\Psi^{(0)}_{0}(x) = 0 .$$
For $ \lambda > \sqrt{-d}\quad ( \lambda < \sqrt{-d} ) $
it occurs when $ \alpha = \mp\sqrt{-d} $ and 
$ \Psi^{(0)}_{0}(x) = \Psi_{2, 1}(x) $ which entails the equation:
\be
f'(x) = -2f^{2}(x) - 2\lambda xf(x) \mp \sqrt{-d} \label{equ}
\ee
with asymptotic behavior $ f(x) \sim \mp \sqrt{-d}/2\lambda x . $
The spacing between the two ground states 
$ \Psi^{(0)}_{0}(x) $ and $ \Psi^{(0)}_{-}(x) $ is:
$ \Delta E = 2\lambda \mp 2\sqrt{-d} .$

e) \, The doublet representation 
$ \bigl( \Psi^{(0)}_{0}(x), \,\, a^{+} \Psi^{(0)}_{0}(x) \bigr) $
of the spectrum generating algebra (\ref{shape}), (\ref{alg2}) is
built on solutions of the equation:
$$
(a^{+})^{2}\Psi^{(0)}_{0}(x) = a^{-} \Psi^{(0)}_{0}(x) = 0 .
$$
It may hold when $ \alpha = 2\lambda + \sqrt{-d}$ for arbitrary
positive value of $ \lambda $ and when $ \alpha = 2\lambda - \sqrt{-d}$ 
for $ \lambda > \sqrt{-d}$. It is equivalent to 
$ a^{+}\Psi^{(0)}_{0}(x) = \Psi_{1, 2}(x) ,$ which is satisfied
when $ f(x) $ obeys the following equation:
\ba
& &8\lambda f^{2}(x)\bigl( f'(x) + 2f^{2}(x) + 2\lambda x f(x) - 
2\lambda \mp \sqrt{-d} \bigr) =
\bigl( f'(x) + 2f^{2}(x) +  \nonumber\\ 
& &2\lambda x f(x) \mp \sqrt{-d} \bigr)\cdot
\biggl[\, \bigl( f'(x) + 2f^{2}(x) + 2\lambda x f(x) -2\lambda 
\mp \sqrt{-d} \bigr) \nonumber\\
& &\bigl( -f'(x) + 2f^{2}(x) + 2\lambda x f(x) -2\lambda 
\mp \sqrt{-d} \bigr) - 4\lambda (\lambda \pm \sqrt{-d}) \biggr] .
\label{*}
\ea
One can show that all solutions of this equation fulfill the Painleve-IV
equation (\ref{painleve}). These solutions have the asymptotics 
$ f(x) \sim \pm \sqrt{-d}/2\lambda x $ and cannot have any (pole)
singularity. The spectrum consists of a doublet $ (0,\,\,2\lambda) $
and infinite sequence 
$ E_{n}=\pm2\sqrt{-d} + 2(n+2)\lambda , \,\, n=0,1,2... $

It may seem to be possible to obtain a higher representation
of spectrum generating algebra with three or more levels in the band.
However for higher order Darboux transformations it is known
\cite{acdi1}  that the zero mode of the third order
operator $ a^{-} $ cannot have more than two nodes, {\it i.e.} it
cannot be higher in energy than the second excited level of system. From 
our analysis it follows that states forming the finite
representation lie below the first state of the infinite sequence of levels,
which is obviously a zero mode of $ a^{-} .$ Therefore this finite 
representation may include only the ground state and, possibly, the 
first excitation. For shape-invariance of $ N $-th order
the maximal dimension of finite representation may be $ (N-1) .$

\section*{\large\bf 4. Conclusions.}
\hspace*{3ex}As conclusive remarks let us now emphasize the novel features of
our approach.
We would like to point out what in our opinion is new in respect to
\cite{veselov}. First, we introduced in the dressing chain 
(in their language) irreducible transformations which correspond
to quantum mechanical systems with only one sequence of levels.
Second, we discovered that in suitable circumstances one 
sequence can be truncated because the higher order creation
operators $a^+$ can have zero modes. Third, we have given  
attention to the asymptotic properties of the Painleve-IV equation
(\ref{painleve}) in connection with discussion of the normalizability
of eigenfunctions. In particular, 
for the construction of
potentials which do not have 
subleading oscillatory behavior,
we revealed the importance of
the special asymptotics $g(y)\sim \sqrt{b}/2y$ which can be implemented
\cite{kapaev}
at both $\pm\infty$ or only alternatively at $+\infty$ or at $-\infty .$
We have found that a particular class of solutions of a Painleve-IV
equation can be obtained as solutions of another, first order, equation
(\ref{*}), which does not admit singularities.
 
Concerning \cite{sukhatme}, we differ because implementation of higher order 
shape-invariance in our scheme does not require to preserve
standard shape-invariance of first order step by step. The spectra
that we can generate appear to be more rich: possibility of having 
three, two, one sequences; possible truncation of one sequence;
non-oscillatory asymptotic behavior of the potential.  

In respect to \cite{hussin} our approach gives more importance to
the distinction between reducible and irreducible second order 
transformation and to potentials which generate sequences of levels spaced
by $2\lambda$ but do not in general have the 
asymptotics behavior $\lambda^2x^2.$

\section*{\large\bf\quad Acknowledgments.}
\hspace*{3ex}
A. A. is grateful to  L.M.Nieto and J.Negro for  discussions
of particular realizations of shape-invariance and for hospitality
in the University of Valladolid.\\
M. I. thanks A.Kapaev for very useful and
exhaustive information
concerning Painleve-IV asymptotic properties. In addition, he would like
to thank INFN and University of Bologna for warm hospitality.
This work was partially supported by RFBR (grant 99-01-00736). 
A.A. is  supported by the grant of Generalitat de Catalunya.

\end{document}